\documentclass{appolb}
\usepackage{epsfig}
\begin{document}
% \eqsec  % uncomment this line to get equations numbered by (sec.num)
\title{MESON2000 Conference Summary {\it lite}
\thanks{Presented at MESON2000 (Cracow, 19-23 May 2000)}
% you can use '\\' to break lines
\author{T.Barnes
\address{Physics Division,  
Oak Ridge National Laboratory\\
Oak Ridge, TN 37831-6373, USA  \\
Department of Physics and Astronomy,
University of Tennessee\\
Knoxville, TN 37996-1501, USA \\
Institut f\"ur Theoretische Kernphysik
der Universit\"at Bonn \\
Nu\ss allee 14-16,
D-53115 Bonn, Germany\\
Institut f\"ur Kernphysik, 
Forschungszentrum J\"ulich\\
D-52425 J\"ulich, Germany\\ }     
}
}
\maketitle
\begin{abstract}
This short contribution is a {\it lite}
MESON2000 conference summary.
As appropriate for the 600th anniversary of the Jagellonian University,
it begins with a
brief summary of the last 600 years of European history and its place
in hadron physics. 
Next a  
``physicist chirality'' order parameter ${\cal PC}$ is introduced. 
When applied to MESON2000
plenary speakers
this  
order parameter
illustrates the separation of
hadron physicists into disjoint communities.
The 
individual plenary talks in MESON2000 are next sorted according to the
subconference associated with each of the 36 plenary speakers. 
Finally, I conclude with a previously unreported 
Feynman story regarding the use of models in hadron physics.
\end{abstract}
  
\section{Hadrons and the Last 600 Years of European History}

I was very happy to accept an invitation from H.Machner to give the summary
talk for MESON2000. However I was quickly surprised and somewhat dismayed
by many of the plenary talks, which discussed  
meson physics from viewpoints 
with which I was only vaguely familiar.
A second theme of this meeting, beginning with Prof. Jarczek's welcoming
talk, was the 600th anniversary of the refounding of the 
Jagellonian University, 
and its place in
Polish and European history. 
Late one night, while puzzling over how to reduce this rather 
broad meeting to a few remarks, the
random thought occurred to me 
that it would 
be easier to summarize the 
last 600 years of European history than to summarize
this conference.
And, like many wild ideas, this one would not 
go away. So, please bear with me through Fig.1.

\begin{figure}[h]
\label{fig_1}
\epsfxsize=5truein\epsffile{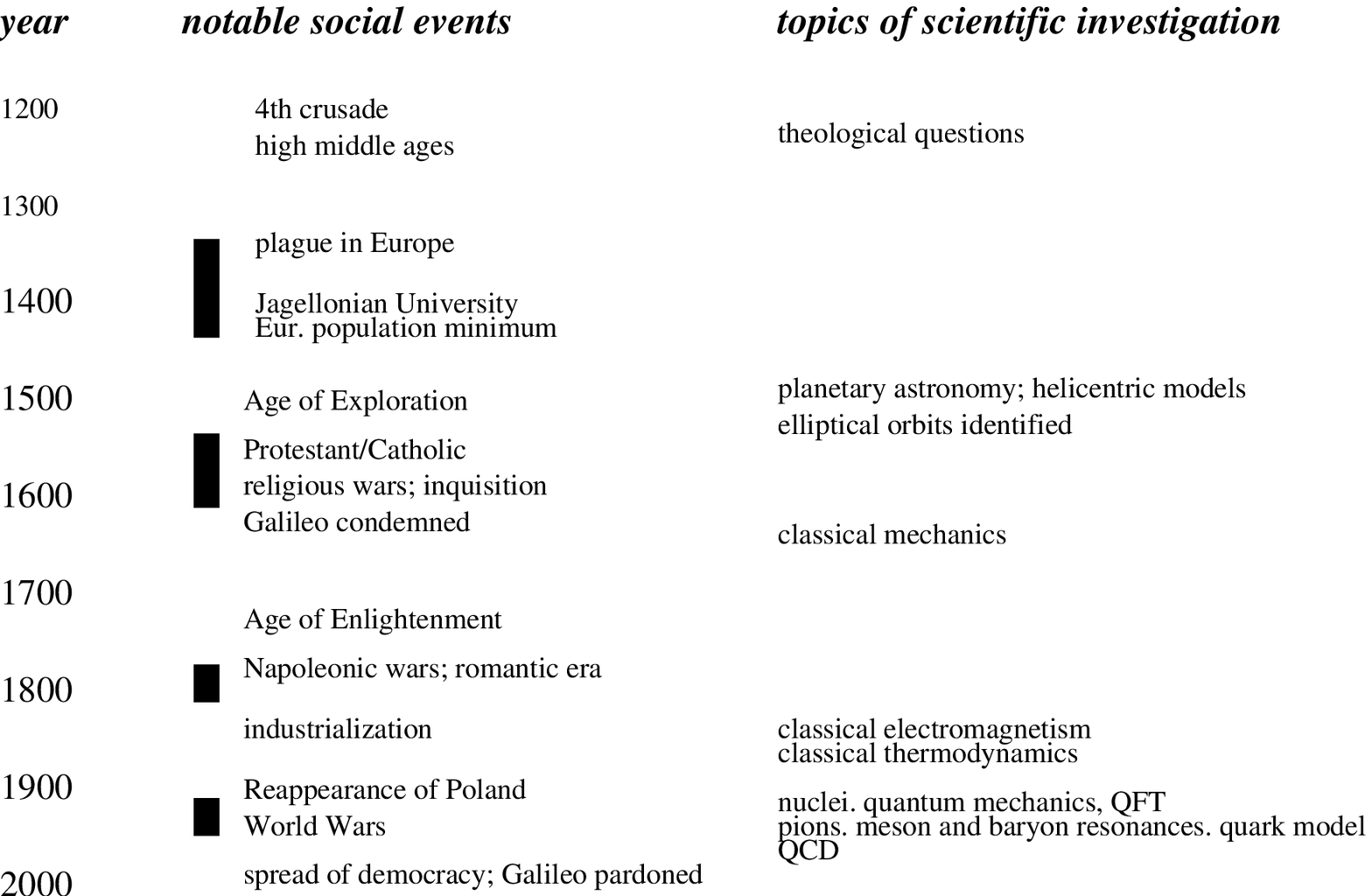}
{\vskip 0.3cm
Fig.1. Hadrons and the last 600 years of European history.}
\end{figure}
One can learn several interesting things from Fig.1.
The first is that, in physics at least, we have made
considerable progress.
At the beginning of this timeline 
scholars were mainly concerned with theological 
considerations that were not amenable to experimental confirmation. 
With the reestablishment of European universities 
and recovery of 
classical texts, physicists again turned to the study of astronomy,
which led to a precise formulation of classical mechanics in the 
17th century. 
The copy of {\it De Revolutionibus Orbium Coelestium} (Copernicus)
which many of us have seen here in Cracow is a stirring reminder of the
work of our antecedents in this field.
To my mind 
European
culture peaked in the 18th century 
Age of Enlightenment, both in the development
of scientific attitudes and socially, in fields
as disparate as government (Jefferson), history (Gibbon), literature (Goethe)
and music 
(Mozart).
With the emergence of the romantic movement in the 19th century, a general
decline was evident; theories of government became more radical, 
nationalism became fashionable in Europe, 
and music and literature increasingly
reflected the social problems of the age. To quote Wittgenstein
regarding music, 
as early as Brahms 
`I can begin to hear the sound of machinery.'
\cite{Witz}.
Physics nonetheless
continued dramatic advances,
due in part to developments in this machinery and in
mathematics, 
and the 19th century saw major achievements in the establishment of
the theories of electromagnetism and classical
thermodynamics. 
Finally, in the 20th century Europe entered very troubled 
times indeed, from which we are only now emerging.
One should note that good things can arise even in times of 
adversity, such as 
the reappearance of an independent Poland in 1919. 
Physics also went through crises at the beginning of the 20th century, but we
must all agree that the resulting quantum physics and 
relativity are two of the most exciting and fascinating developments in 
the field. 

Against this 600 year historical background the development of hadron physics
has been strikingly rapid. The identification of the compact atomic nucleus,
the home of most terrestrial hadrons, 
was due to Rutherford in 1911.  
The identification of the positively charged 
proton, the first known hadron, can also be dated to about 1911.
The first meson to be identified was the pion,
found by Lattes {\it et al.} 
in 1947 (in cosmic rays), and it had been anticipated by Yukawa as the carrier
of the strong nuclear force. The familiar light mesons $K$, $\rho$, $\omega$
etc were found in the late 1950s to early 1960s, and the identification of these
and the light baryons
suggested the quark model to Zweig, Gell-Mann and Ne'emann in about 1963.
The identification of QCD as the theory of the strong interaction, in 1973, was
due mainly to its property of asymptotic freedom, which had been observed 
at SLAC in the late 1960s. 
The crucial confining property of QCD was at the time regarded as an unproven
conjecture, and is still poorly understood.
The mid to late 1970s saw the experimental establishment
of the charm and beauty families of hadrons, the first searches for glueballs, 
and the development of new 
theoretical techniques such as LGT. The remarkable QCD predictions of
glueballs and exotic mesons have taken longer to test experimentally, and 
the more widely accepted experimental candidates
for these states 
were identified in the middle 1990s. 
This short time
scale is most reassuring; in Fig.1 we can see that almost all the 
progress in strong interaction physics 
has been made in the last 10\% of the timeline, and the study of QCD itself
occupies only the final 5\%.

And finally, as if to close the circle, at the end of the millennium many
theoreticians have again turned to theological
speculations which are not amenable to experimental confirmation.

\section{The Many Phases of the Meson Community}

After the first few plenary talks I was confirmed in my suspicion that
hadron physics, even meson physics, is a very broad field with 
clearly identifiable communities that have little overlap. Since much of
the work in contributing to a new field involves learning the field's 
terminology or ``jargon'', one can identify the different 
communities
by the rate of recurrence of characteristic 
words or expressions in
research papers or presentations at conferences. 

Two obvious communities in nonperturbative QCD 
are the effective lagrangian / chiral symmetry 
specialists (to whom the pion is the most interesting
meson) and the
hadron spectroscopists (to whom it is not).
As a test of the idea of separating these groups by their use of language,
I invented an order parameter which I call ``physicist
chirality'' $({\cal PC})$ to distinguish them. 
${\cal PC}$ is defined by the number of times a plenary speaker used
the word ``chiral'' compared to the exotica words ``glueball'', ``hybrid'' or
``exotic'',
\begin{equation}
{\cal PC} = { N_{chiral} - N_{exotica} \over N_{chiral} + N_{exotica} }
\ .
\end{equation}
This quantity is $-1$ for a purely 
exotic physicist and $+1$ for a purely chiral 
physicist. I had expected to find an interesting distribution of ${\cal PC}$
in this meeting, so I applied it to the first 18 plenary speakers. The result,
shown in Fig.2, 
is rather disturbing. One sees clear evidence of 
``phase separation'' in the presence of 
two almost completely disjoint communities in meson physics!

\begin{figure}[h]
\label{fig_2}
\hskip1.5cm
\epsfxsize=3truein\epsffile{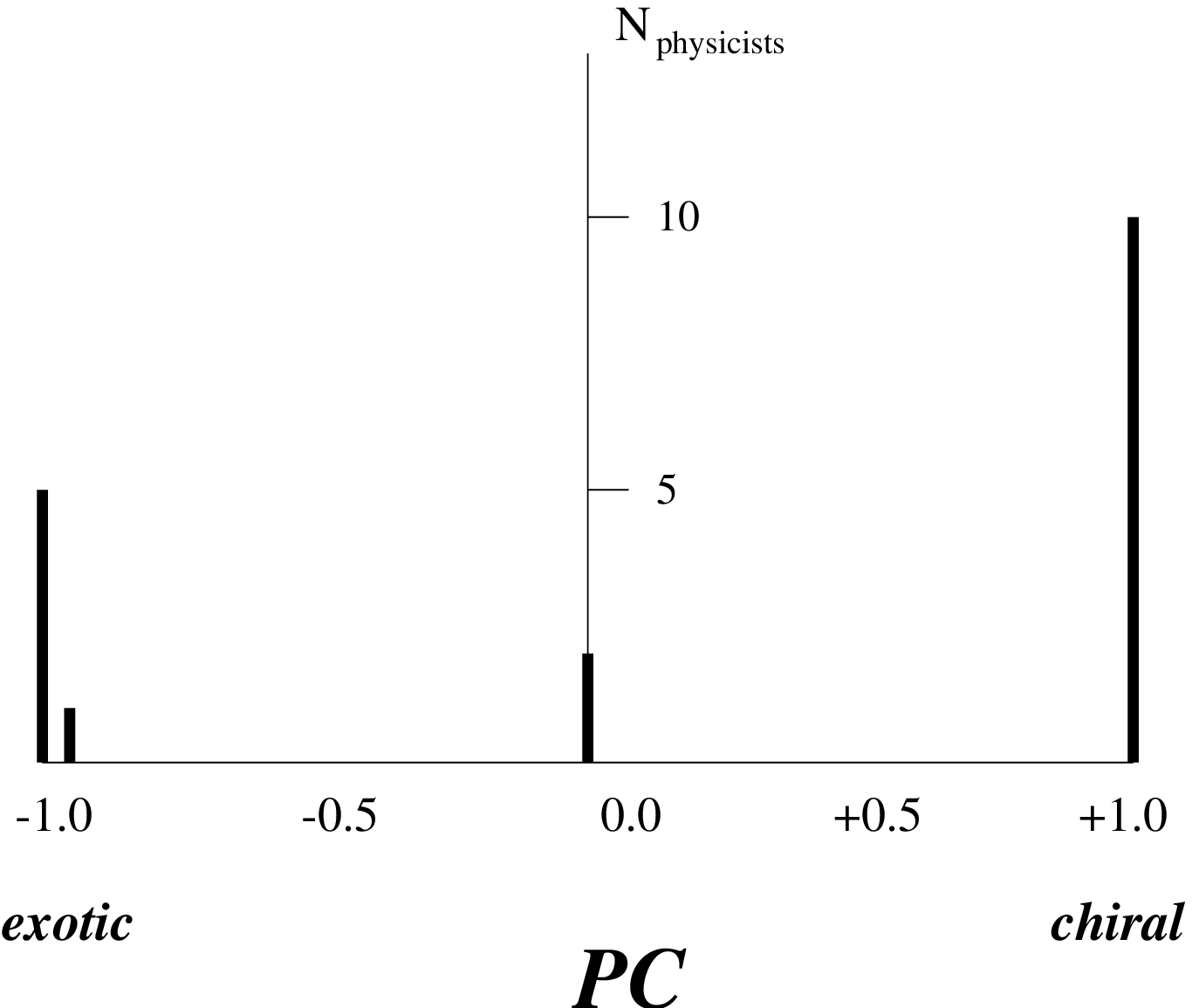}
{\vskip 0.5cm
\hskip1cm
Fig.2. Phase separation in the hadron physics community.}
\end{figure}
For those who are keeping score, the two extreme cases on an absolute
scale were 
W.Weise (39 uses of chiral)
and
A.Szczepaniak (over 29 uses of exotic).
There were also 0/0 scores (recorded at 0), 
which suggested additional phases; eventually 
I identified five subconferences at MESON2000, which are summarized
in the following section.

\section{The index of plenary speakers and their conferences}

\subsection{The Condensed Matter Hadron Conference}

This subdiscipline of meson physics considers the pion to be the most 
interesting of mesons because it is the lowest-lying excitation, 
in condensed matter jargon the 
``gap mode''. 
In this field one 
discusses hadrons in terms of the $\langle q\bar q \rangle$ ``order 
parameter'', writes effective lagrangians for pions and nucleons, and 
derives the resulting low-energy scattering amplitudes and 
in-medium properties. 
Apparently part of the game is to borrow as much of the
condensed matter viewpoint and jargon as possible, 
and use it in a hadronic context.
(I am not being entirely frivolous about what appears to
me to be an exercise in borrowing 
jargon, since I actually work in
condensed matter physics \cite{magmag}.)

Our first plenary speakers 
Weise, Thomas and (rather surprisingly for an 
impartial experimentalist) 
Nefkens identified themselves as 
belonging to this community through their
frequent references to chiral symmetry and order parameters. 
Mosel, Senger, Grioni, Cassing and Oset 
also participated in this subconference, 
which was
largely 
concerned with prospects for seeing mass shifts
e.g. of kaons, nucleons and vector mesons in medium, 
and in-medium corrections to other processes, such as
$\pi\pi$ scattering. 
In these talks one could often hear loan words from
condensed matter, extending in extreme cases to
``spectral functions'' 
and even ``good quasiparticle'', which is
a rather imprecise notion even in 
quantum magnetism.

Throughout I confess to having a feeling that we need one
very clear, unambiguous experimental observation of a 
hadronic in-medium mass shift in a 
relatively narrow state,
such as the $\omega$, before this work on mass shifts can be regarded as
supported by experiment; the inclusive distribution of 
dilepton mass pairs,
composed of many hypothetical broad contributions, 
seems less than convincing.
Cassing noted in particular that there are several 
interesting prospects for observing 
$\omega$ mass shifts.

Experiment has not contributed much to this field ``in-vacuum" recently,
largely because the favored topics of low energy $\pi\pi$ and 
$\pi N$ scattering were studied long ago, {\it albeit} not with
especially good statistics in $\pi\pi$. 
At this meeting however we heard plans 
for two new relevant measurements. One is
a determination of the $a_2 - a_0$ $\pi\pi$
scattering length difference using $\pi^+\pi^-$ atoms, discussed by
Gianotti. There are dispersion relations known as the
Roy equations that purportedly constrain these quantities accurately, 
so this will be a useful measurement.
The second (at DAPHNE, discussed here by Lauss) 
is a measurement of $\bar K N$ 
$a_0$ and $a_1$ scattering lengths, similarly using 
radiative transitions in the hadronic atom.

\subsection{The Good Auld Hadrons Conference}

This second subconference was concerned with ``good auld hadrons'', 
by which
I mean low energy reactions and the spectroscopy 
of reasonably well established
quark model states. There was a large experimental 
component from the various
few-GeV facilities that are studying these processes. 
Meson production
near threshold, especially as accessible at COSY, 
was discussed by Speth. The interesting topics here are the
various possible mechanisms for $\pi$ and $\eta$ 
production and the recurring
question of the physical size of the $qqq$ ``quark core'' in a baryon.
Haberzettl gave a clear review of the complications of
electroproduction amplitudes, in particular in the production of
associated strangeness
at CEBAF. Filippi told us about OZI violation in
$P\bar P$ annihilation, especially in the final 
state $\phi\pi$, 
and discussed
ways of distinguishing between intrinsic strangeness and 
rescattering effects such as
$K^*\bar K\to \phi\pi$. Moskal discussed 
$PP\to PPf$ at COSY-11, where $f=\pi,\eta,\eta'$ 
and $K^+K^-$, and noted the
importance of ISI and the possibility of measuring the 
$P\eta'$ scattering 
length. Salaburn discussed $PP\to PPf$ at 
DISTO, where one can also 
study $f=\rho,\omega,\phi$. He noted that one may test 
hidden-$s\bar s$
components using polarized $PP\to PP\phi$, and also that
the $PP\to P K^+\Lambda$ data supports a simple $K$-exchange picture. 
Str\"oyer
discussed baryon resonance production and 
decays, and noted 
that such a programme at COSY using $P$ beams would be a 
useful complement to 
the various photon facilities now planned or in operation.
Niskanen discussed the origin of 
isospin violation effects
in nucleon-nucleon scattering and suggested channels 
in which these effects may be
largest. 
Klimala 
considered meson production in $PD$ collisions
and discussed 
how one might test models such as intermediate $\Delta$ production.
Eyrich discussed prospects
for strangeness production at
COSY using TOF, including $K\Lambda$ (current), 
$N^*$s (possible in future),
and the very 
interesting but long neglected
$Z^*$ exotic flavor channels (in particular the sector $uu\bar d\bar d s$).
Kupse summarized future plans for 
CELSIUS/WASA, which include
rare $\pi^o$ and $\eta$ decays, including the processes
$\eta\to \pi^o\pi^+\pi^-$ (I violating, of interest in $\chi$PT)
and
$\eta\to \gamma\pi^+\pi^-$ (for which VMD and  $\chi$PT give rather
different predictions). Finally, Braccini reviewed the
very interesting results on $\gamma\gamma$ couplings 
of $n\bar n$, $s\bar s$ 
and $c\bar c$ resonances which have come from 
LEP and Cornell 
recently. These results include observations of the 
possible radial 
excitations $\eta(1440)$ and $a_2'(1752)$, and 
in $\gamma\gamma^*$ evidence
for the light axials $f_1(1285)$ and $f_1(1440)$ and 
the $\eta_c$, which
is apparently produced through $J/\psi$ vector dominance.

\subsection{The Exotica Conference}

The subject of non-$q\bar q$ mesons, the so-called ``exotica'', has seen
exciting developments of late, with the announcement of glueball and
spin-parity exotic candidates. This, I admit, 
was the conference I attended.

Barnes first gave a review of exotic mesons 
(exotic meaning 
having quantum numbers forbidden to conventional $q\bar q$ states). 
We now
have two light exotic candidates, the
$\pi_1(1400)$ (BNL, Crystal Barrel, VES)
and 
$\pi_1(1600)$ (BNL, VES). Unfortunately 
it may be too soon to celebrate, 
since LGT predicts that the $1^{-+}$ exotic level 
should lie at about 2.0 GeV,
much higher than reported.
Klempt discussed glueballs, 
in particular the various states in 
the $0^{++}$ sector. These include the 
Crystal Barrel candidate $f_0(1500)$, the
possibility of a single very broad scalar 
``der Rote Drache'', and various models 
of scalar mixing and decays.
Szczepaniak
summarized the CEBAF HallD project, which 
is a proposed high-statistics meson photoproduction 
experiment 
for the study of light exotic and $s\bar s$ meson spectroscopy. 
Willutski
reviewed results from 
BNL E852, including
evidence for the $\pi_1(1590)$, and in $\omega\eta$ for a
$1^{+-}$ $h_1(1590)$ and a
$1^{--}$ $\omega(1650)$. Note that the $h_1(1590)$ 
is considerably lighter
than expected by Godfrey and Isgur for a 2P state.
Close reviewed his very interesting work on the exchanged angular
quantum numbers 
and $Q_i^2$ 
dependences in diffractive meson production, which can
be used as a ``discriminator'' between $q\bar q$ and $G$ candidates.
Clearly something very important about diffraction 
has been discovered here,
although just what is as yet unclear.
Peters restricted himself to the ``past, present and future'' of meson 
spectroscopy, including the 10th anniversary of the Crystal Ball
$3\pi^o$ Dalitz plot, developments in exotics, the importance of high
statistics, complications in analyses, scalars, D decays, ... and 
future facilities. Finally, Stefanski 
reviewed results from
the charm photoproduction experiment E791 at Fermilab, 
and noted that the
$3\pi$ Dalitz plots from $D^+$ and $D_s^+$ show 
evidence for strong
isobar contributions, including $\rho\pi$, 
$f_0\pi$
$f_2\pi$ and
$\sigma\pi$. The interesting evidence of FSI effects 
in the complex relative 
phases of these states was also noted.

\subsection{The HEP Conference}

Subconference 4 (on HEP) was the shortest, with just two plenary contributions.
These contributions could 
be identified by the
fact that the hadrons were clearly considered non-essential 
complications
to the interesting physics. The first HEP contribution was by
Sciaba, who summarized the status of the search for 
$B_s^o \leftrightarrow \bar B_s^o$ mixing. This has not yet 
been observed, but 
the limits are now rather close to theoretical expectations, 
``watch this spot''. 
Next, Fleischer reviewed the general
subject of CP violation in B decays in impressive detail, and suggested 
several final states which may be of
interest in future experimental studies. 

\subsection{The Photon-Hadron Conference}

The final subconference I identified, with six plenary contributions, 
was on
photon-hadron interactions. (Braccini's two-photon talk might also be 
listed here as a seventh contribution.)
The first contribution was by
Levi Sandri,
who reviewed the baryon resonance 
program at GRAAL, and
noted some interesting results, such as the fact 
that the 
$D_{13}(1520)$ 
$A_{3/2}/ A_{1/2}$ 
ratio 
does not agree well with 
Capstick and Isgur's quark model
predictions. 
Bruncko
reviewed DESY results on vector photoproduction of
$\rho$, $\omega$, $\phi$, $\psi$, $\psi'$ and $\Upsilon$. A remarkable
``universal curve'' of electroproduction cross sections 
versus $Q^2$ was shown
for these states. 
Steffens reviewed polarized deep inelastic scattering at HERMES,
especially the ``semi-inclusive'' processes
$\gamma^*P \to  \rho, \omega, \phi, \psi, \Upsilon +X$, tests of SCHC, 
production mechanisms and parton distributions.
Arends reviewed the status of the 
DHG sum rule and concluded that there is no indication
of disagreement with experiment. 
Nikolaev discussed diffractive electroproduction of vector mesons and
the interesting possibility of distinguishing 2S from D states through their
different 
$Q^2$ dependences. 
And finally, Muccifora discussed  charged and neutral $\pi$
electroproduction at HERMES, which can be used to test the $Q^2$ evolution of 
fragmentation functions. These results included surprising evidence of 
possible 
isospin symmetry violation above $z=0.7$.
 
\section{Personal Favorites}

Although there were many interesting results presented at the meeting,
I would like to take advantage of my r\^ole as summary speaker 
to cite what 
seemed to me personally to be the single most remarkable new experimental
and theoretical results. 

In experiment: The ``universal curve'' for the
$Q^2$ dependence of vector meson electroproduction appears to be a very 
suggestive observation, and presumably tells us something very general about
hadron 
electromagnetic couplings. Does this establish a vector dominance picture
over direct photon-quark pQCD amplitudes? If so, most quark model
calculations of resonance photoproduction and electroproduction 
amplitudes may be inaccurate! The question of just what this 
result teaches us should clearly be pursued. 

In theory: Close has found remarkably
simple and accurate 
results for diffractive scattering amplitudes, using an almost
conserved vector coupling model; this is telling us 
{\it something} profound about
the long standing issue of just what the ``pomeron'' is at the 
quark-gluon level. As with many interesting discoveries, it is not yet clear
what these results mean, but they suggest that progress in this long-standing
question may now be possible.

\section{Feynman Story}

At this conference, 
the question of the limits of usefulness
of various models must have occurred to many of the attendees.
This story gives some indication as to Feynman's attitude to the use
of models 
in 
hadron physics. 

In about 1974 as a new Caltech graduate student I was looking for an 
interesting thesis topic. This was an exciting period with many new 
developments in physics, such as
supersymmetry, string theory, the parton model, 
non-Abelian gauge theories, compact objects in
astrophysics and so forth. Although I was initially interested in
rather formal problems in quantum gravity, the very practical and
skeptical
research atmosphere at Caltech strongly encouraged graduate students
to study topics that led to direct comparison with
experiment. Since QCD had just been proposed, and a
group at
MIT had just published their first paper on the ``bag model'' which
showed that one could derive many experimentally observable properties of
light hadrons 
quite simply using quark and gluon ``cavity resonator'' modes,
I began work on this model and suggested it as a thesis
topic to my advisor Jon Mathews. Since Mathews was a rather pure mathematical
physicist, he was unenthusiastic. He suggested however
that I talk to Feynman about this work, 
since Feynman had heard about the model
at a meeting and had since worked out many of its predictions for hadrons 
himself. I found that Feynman, unlike Mathews, 
was very interested in
and excited by what could be derived in this simple model; so much so that I
rather courageously 
asked him if he would tell Mathews that the study of this model 
was a suitable 
thesis topic. 
The resulting transition from initial to final states
is shown in Fig.3 below (this is a Feynman diagram in which Feynman 
actually appears).

\begin{figure}[h]
\label{fig_3}
\hskip1.5cm
\epsfxsize=3.2truein\epsffile{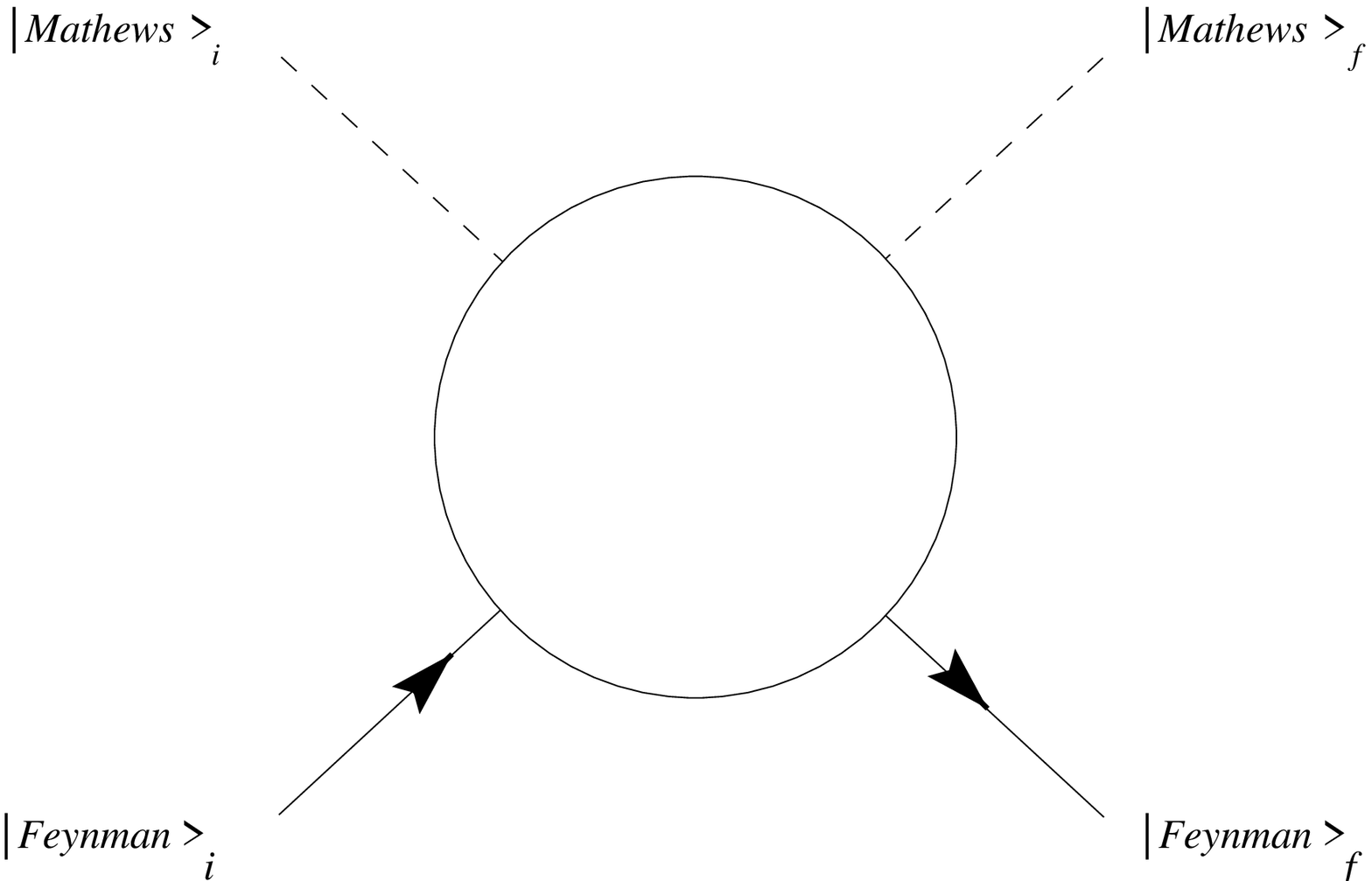}
{\vskip 0.3cm 
\hskip1.5cm
Fig.3. A Mathews-Feynman scattering diagram.}
\end{figure}

%\vskip 0.5cm

As usual, precisely what took place
within the circle is unknown, but much can be inferred from 
the initial and final states.
In this case, in the initial state 
\ $|Mathews\rangle_i$ \
the MIT bag model was {\it not} a suitable Ph.D. thesis topic, 
and in 
the final state
$|Mathews\rangle_f$ 
it {\it was} a suitable topic. Although the details of the
interaction were not observed, 
\ $|Mathews\rangle_f$\  
made statements to the effect that {\it the study of models 
is useful in hadron physics because
one can abstract model-independent features}. I presume that this is the
justification Feynman gave to Mathews. 
This has indeed 
proven to be the case for the bag model, 
since it was the first to predict a light $J^{PC}=1^{-+}$ exotic meson;
this exotic 
has been found by all subsequent approaches, including LGT. We now
have two experimental exotic meson candidates with these quantum numbers,
which were discussed in detail at this meeting, and the
general topic of ``exotica'' 
is now widely 
considered the most interesting subject in light hadron spectroscopy. 

In summary, each model is wrong in detail, but they may nonetheless contain
some common physical truth.

\section{Acknowledgements}

It is a great pleasure to thank the organisers 
for their kind invitation 
to summarize the many conferences of Meson 2000.
This research was supported in part by the 
DOE Division of Nuclear Physics,
at ORNL,
managed by UT-Battelle, LLC, for the US Department of Energy
under Contract No. DE-AC05-00OR22725, and by 
the Deutsche Forschungsgemeinschaft DFG
at the University of Bonn and the Forschungszentrum J\"ulich
under contract Bo
56/153-1.

\end{document}